\documentclass[prl,superscriptaddress, preprintnumbers ,twocolumn,showpacs,amsmath,amssymb]{revtex4}
\usepackage{amsmath}
\usepackage{amssymb}
\usepackage{amsthm}
\usepackage{amsfonts}
\usepackage{enumerate}
\usepackage{latexsym}
\usepackage{graphicx}

\begin{document}

\title{Localization and Kosterlitz-Thouless Transition in  Disordered Graphene}
\author{Yan-Yang Zhang}
 \affiliation{ Department of Physics, Purdue
University, West Lafayette, Indiana 47907, USA}
\author{Jiangping
Hu} \affiliation{ Department of Physics, Purdue University, West
Lafayette, Indiana 47907, USA}
 \author{ B.A. Bernevig}
 \affiliation{
Princeton Center for Theoretical Science, Jadwin Hall, Princeton
University, Princeton, NJ 08544}
\author{X.R. Wang}
  \affiliation{ Physics Department,
The Hong Kong University of Science and Technology, Clear Water Bay,
Hong Kong SAR, China.}
\author{X.C Xie}
 \affiliation{  Institute of Physics, Chinese Academy of
Sciences, Beijing 100080, China.} \affiliation{Department of
Physics, Oklahoma State University, Stillwater, Oklahoma 74078, USA}
\author{W.M Liu}
 \affiliation{  Institute of Physics, Chinese Academy of
Sciences, Beijing 100080, China.}
\date{\today}

\begin{abstract}
We investigate disordered  graphene  with strong long-range
impurities. Contrary to the common belief that delocalization should
persist in such a system against any disorder, as the system is
expected to be equivalent to a disordered two-dimensional Dirac
Fermionic system, we find that states near the Dirac points are
localized for sufficiently strong disorder and the transition
between the localized and delocalized states is of
Kosterlitz-Thouless type. Our results show that the transition
originates from bounding and unbounding of local current vortices.
\end{abstract}


\maketitle It is well-known that the electronic spectrum of graphene
can be approximately described by relativistic Dirac
Femions\cite{Wallace47,Nov05}. This is due to the linear dispersion
relation at low energies near two valleys associated to two
inequivalent points $\mathbf{K}$ and $\mathbf{K}'$ at the corner of
the Brillouin zone\cite{Kane05}. The relativistic dispersion gives
rise to several remarkable phenomena. Unlike non-relativistic
Schr\"{o}dinger fermions in two dimensions \cite{Abrahams79}, Dirac
fermions cannot be trapped by a barrier due to the Klein paradox, a
property of relativistic quantum mechanics\cite{Katsnelson06}.
Theories based on the two-dimensional (2D) single flavor Dirac
Hamiltonian also predict that Dirac fermions cannot be localized by
disorder\cite{Ando98,Ziegler98,Bardarson07,Nomura07}.

The great majority of experimental and theoretical studies of
graphene \cite{Nov05,Huard07} has focused on the effect of the
relativistic electronic dispersion on different phenomena such as
Landau level structure or quantum Hall ferromagnetism. However, the
validity of single flavor Dirac fermion picture for disordered
graphene is only approximate and relies on two premisses: (1) The
spatial range of the impurities is long enough to avoid inter-valley
scattering\cite{Ando98} (for short-range impurities, strong
inter-valley scattering  can lead to
localization\cite{Ando98,Atland06,Xiong07}); (2)  Even in a
disordered graphene with  long-range impurities that completely
suppress inter-valley scattering, a single valley Dirac Hamiltonian
is only valid  when  weak impurities are considered. Since the
approximated linear relativistic dispersion is valid near the Dirac
valley points, $\mathbf{K}$ and $\mathbf{K}'$, the approximation
cannot be carried out in a region with a strong impurity where the
deviation from the Dirac point is large enough so that higher order
corrections to the energy spectrum become relevant\cite{Ando98}.
Therefore, the application of single valley Dirac Hamiltonian to
disordered graphene is limited to weak long-range impurities.
Indeed, localized states in disordered graphene near Dirac points
have been observed experimentally \cite{Adam08} and
numerically\cite{Pereira06,Amini08}. All the above beg the physical
question: how does graphene behave in the presence of strong
long-range impurities?

In this letter we investigate several novel phenomena induced by
disorder with strong long-range impurities in graphene. We calculate
the scaling properties of disordered graphene in the framework of a
tight-binding model and finite-size scaling. Instead of
delocalization we find that, in the presence of strong long-range
impurities, states near the Dirac points are localized.
Localization arises from enhanced backscattering due to the
deviation from linear dispersion in the strong impurity regime.
We show that there is a metal-insulator transition (MIT) as a
function of the disorder strength and chemical potential. On the
delocalized (metallic) side, the conductance is independent of the
system size, which is a characteristic of  the Kosterlitz-Thouless
(K-T) \cite{Kosterlitz73} type transition in conventional 2D systems
with random magnetic field \cite{Xie98,Liu99} or correlated disorder
\cite{Liu99B}.  We verify the Kosterlitz-Thouless transition nature
of the MIT by explicitly identifying the  bounding and unbounding
vortex-anti-vortex local currents in the system.

The $\pi$ electrons in graphene are described by the tight binding
Hamiltonian (TBH)
\begin{equation}
H=\sum_{i}V_{i}c_{i}^{\dagger }c_{i}+t\sum_{\langle i,j\rangle
}(c_{i}^{\dagger }c_{j}+\text{H.c}),  \label{1}
\end{equation}%
where $c_{i}^{\dag }$ ($c_{i}$) creates (annihilates) an electron on
site $i$ with coordinate $\mathbf{r}_i$,  $t$ ($\sim $2.7eV) is the
hopping integral between the nearest neighbor carbon atoms with
distance $a/\sqrt{3}$ ($a\sim 2.46\AA$ is the lattice constant), and
$V_{i}$ is the potential energy. In the presence of disorder, $V_i$
is the sum of contributions from $N_I$ impurities randomly centered
at $\{\mathbf{r}_m\}$ among $N$ sites
$V_{i}=\sum_{m=1}^{N_I}U_{m}\exp(-|\mathbf{r}_i-\mathbf{r}_m|^2/(2\xi))$,
where $U_{m}$ is randomly distributed within $(-W/2,W/2)$ in units
of $t$. Different random configurations of graphene samples with
same size, $\xi$, $W$ and $n_i\equiv N_I/N$ constitute an ensemble
with definite disorder strength. This model has been widely used in
investigating the transport properties in
graphene\cite{Rycerz07,Waka07,Lewen08}.

At zero temperature, the two terminal dimensionless conductance
$g_L$ of the sample between perfect leads at Fermi energy $ E_{F}$
can be written in terms of Landauer-B\"{u}ttiker formula \cite{Datta}:
\begin{equation}
g_L(E_{F}) = 2\text{Tr}(tt^{\dagger}),\label{3}
\end{equation}
where $t$ is the transmission matrix and the factor 2 accounts for
spin degeneracy. Equation (\ref{3}) can be numerically evaluated by
recursive Green's function method \cite{MacKinnon85} for systems
with rather large size. For the purpose of scaling, the contact
effect should be subtracted from $g_L$ to yield the ``intrinsic
conductance'', $g$, defined as $ 1/g=1/g_L-1/(2N_C)$, where $N_C$ is
the number of propagating channels at Fermi energy $E_F$ and
$1/(2N_C)$ is the contact resistance\cite{Braun1997}. The
conductance $g$ then receives contributions solely from the bulk and
thus has the same scaling property as if it were obtained by the
transfer matrix method \cite{Keith2001}. The scaling
function\cite{Abrahams79,Keith2001}
\begin{equation}
\beta=\frac{d\langle \ln g\rangle}{d\ln L}, \label{7}
\end{equation}%
$\langle \ldots\rangle$ being the average over random ensemble, is
used to determine the localization properties; $\beta<0$ and
$\beta>0$ correspond to the insulator and the metal, respectively.

\begin{figure} [t]
\begin{center}
\includegraphics[bb=15 15 230 288,width=0.4\textwidth]{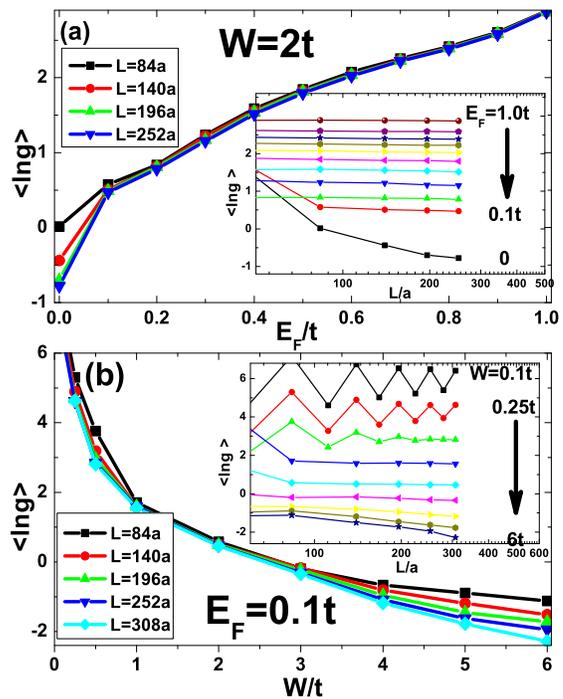}
\end{center}
\caption{(Color online) The scaling of conductance for long range
disorder ($\xi=1.73a$, $n_I=1\%$): (a)$\langle \ln g\rangle$ as
functions of the Fermi energy $E_F$ with fixed disorder strength
$W=2t$(note: the bandwidth is $6t$)); (b) $\langle \ln g\rangle$ as
functions of disorder strength $W$ with fixed Fermi energy
$E_F=0.1t$. The insets are the same data plotted as functions of
size $L$. Each $\langle \ln g\rangle$ is an average over
$100\sim400$ random realizations. } \label{F1}
\end{figure}

\begin{figure} [t]
\begin{center}
\includegraphics[bb=20 20 272 218,width=0.35\textwidth]{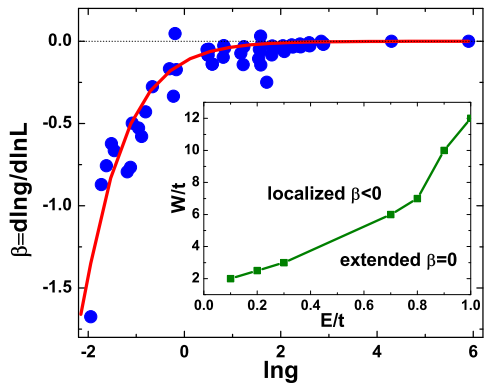}
\end{center}
\caption{(Color online) The scaling function $\beta=\frac{d\langle
\ln g\rangle}{d\ln L}$ obtained from the data in Fig. \ref{F1}. The
inset is the schematic phase diagram for $\xi=1.73a$, $n_I=1\%$.}
\label{F2}
\end{figure}

We plot the size dependence of $\langle \ln g\rangle$ with
$\xi=1.73a$, $n_I=1\%$, for different $E_F$ and $W$ in Fig.
\ref{F1}. The samples are set to be square shaped with length $L$.
Periodic boundary conditions in the transverse direction are adopted
to exclude the edge states of the zigzag edges\cite{Waka07}. The
potential range $\xi$ here is chosen long enough to avoid obvious
inter-valley scattering \cite{Ando98,Waka07}, and the scaling
$\xi/L_x\sim0$ is irrelevant. When $E_F<E_c=0.1t$ (Fig. \ref{F1}
(a)) or $W>W_c=2t$ (Fig. \ref{F1} (b)), $\langle g\rangle$ is
monotonically decreasing with increasing $L$, which means the
wavefunctions are localized. Otherwise, when $E_F>E_c=0.1t$ (Fig.
\ref{F1} (a)) or $W<W_c=2t$ (Fig. \ref{F1} (b)), $\langle \ln
g\rangle$ curves for different sizes merge, suggesting a delocalized
state with finite conductance in the thermodynamic limit. However,
they are not real metals with $\beta>0$. All the states with
$W\in(0,W_c)$ are within the metal-insulator transition (MIT) region
with $\beta=0$. Even in the cases of extremely weak disorder with
$W=0.25t$ and $W=0.1t$ (see the inset of Fig. \ref{F1} (b)), except
for a vanishing even-odd like fluctuation, $\langle\ln g\rangle (L)$
doesn't seem to show a tendency to be increasing nor decreasing. In
Fig. \ref{F2}, the universal $\beta(\ln g)$ is plotted from the same
data in Fig. \ref{F1}, showing a critical conductance $\ln g_c\sim
1$ separating the delocalized states with $\beta=0$ and localized
states with $\beta<0$. This phenomenon corresponds to a
disorder-driven Kosterlitz-Thouless (K-T) type transition that has
been observed in many disordered 2D
systems\cite{Zhang93,Xie98,Liu99,Liu99B}. As can be seen from Fig.
\ref{F1} (a) and the phase diagram (inset of Fig. \ref{F2}), states
in the low energy region are more easily localized.

The existence of localized states near the Dirac point is in
contrast to the belief that Dirac fermions are robust against
localization, especially in the presence of long range impurities
that can effectively prohibit the inter-valley scattering. In order
to gain insight in the nature of the localization transition, we now
turn back to the dispersion structure of realistic graphene. In the
absence of disorder ($V_i\equiv0$), the upper ($+$) and the lower
($-$) bands touch at two Dirac points
$\mathbf{K}=(\frac{2\pi}{3a},\frac{2\pi}{3\sqrt{3}a})$ and
$\mathbf{K}'=(\frac{2\pi}{3a},-\frac{2\pi}{3\sqrt{3}a})$. When
$|E|\leq t$, the dispersion consists of two valleys centered at
Dirac points. Near each Dirac point, e.g. $\mathbf{K}$, the energy
bands can be expanded as\cite{Ando98}
\begin{equation}
E_{\pm}(\mathbf{q})=\pm
\frac{3ta}{2}|\mathbf{q}|\pm\frac{\sqrt{3}ta^2}{8}\sin{(3\alpha(\mathbf{q}))}|\mathbf{q}|^2+O(q^3),
\label{4}
\end{equation}
where $\mathbf{q}\equiv\mathbf{k}-\mathbf{K}$ is the momentum
measured from $\mathbf{K}$ and $\alpha(\mathbf{q})\in[0,2\pi)$ is
the angle of vector $\mathbf{q}$. The first term in the r.h.s of
(\ref{4}) corresponds to the Dirac Hamiltonian, but non-linear terms
will be prominent when $\mathbf{q}$ (or $E$) is increased. Even when
$|E|<t$, the second term in (\ref{4}) can lead to non-trivial
consequences. First, the quadratic dependence on momentum
$|\mathbf{q}|^2$ gives rise to non-vanishing backscattering
probability and thus a tendency to localization. Second, the angular
dependence factor $\sin{(3\alpha(\mathbf{q}))}$ (``trigonal
warping'') breaks the perfect symmetry of the cone-like valley. The
pseudo-time reversal symmetry \cite{Nomura07,Waka07} restricted to
each valley is destroyed. When $|E|>t$, the linear approximation and
double-valley structure collapses completely. Although graphene
cannot be experimentally doped to a bulk Fermi energy far away from
the neutral point (Dirac points) (e.g., $E_F\sim t$), the local
potential of impurities might still be high enough to create
non-Dirac scatterings. The observed localization originates from the
non-Dirac behavior due to higher order corrections to the
dispersion.

\begin{figure} [t]
\begin{center}
\includegraphics[bb=20 20 320 320,width=0.45\textwidth]{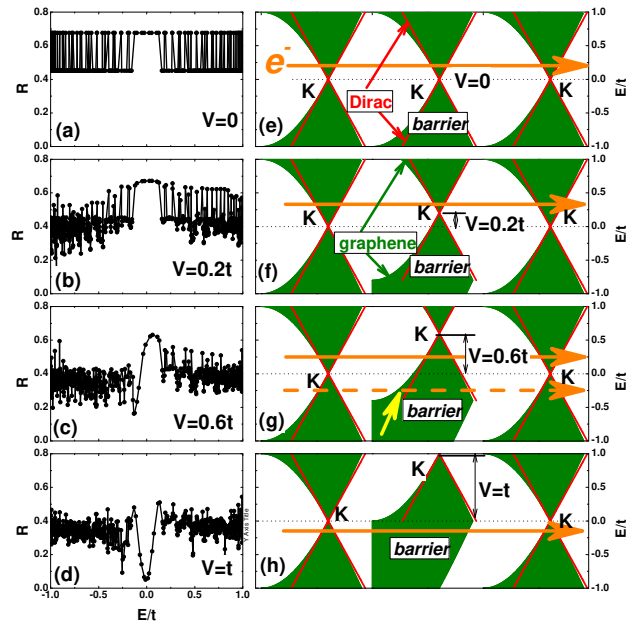}
\end{center}
\caption{(Color online) Left column ((a)$\rightarrow$(d)): The
participation ratio $R$ as functions of energy $E$ for a graphene
with $N=70\times 40$, in the presence of a single impurity at the
center with $\xi=1.73a$ (long range) with different potential height
$V\geq0$. Right column ((e)$\rightarrow$(h)): Schematic diagrams of
scattering process corresponding to their left counterparts. The
electron is injected from the left, scattered by the barrier at the
center and eventually transmitted to the right (thick orange
arrows). The dispersion configurations in these three regions around
$K$ are plotted, where the red lines mark the ideal Dirac dispersion
$E_{\pm}(\textbf{k})=\pm \frac{3ta}{2}|\textbf{k-K}|$ and the olive
part is that for graphene calculated from TBH. Discrepancies between
them at high energy can be clearly seen.} \label{F3}
\end{figure}

To confirm this, let us consider the simplest case of a single long
range impurity in the center of a graphene sheet. If inter-valley
interaction is effectively prohibited and the regime of Klein
tunneling\cite{Katsnelson06} holds, there should be no bound states,
no matter how high the potential barrier is. After diagonalizing the
Hamiltonian for a graphene sheet with $N$ sites, the spatial
extension of eigenstate $|\psi_n\rangle=\sum_{i=1}^N
a_{ni}c_{i}^{\dagger}|0\rangle$ with eigen-energy $E_n$ can be
characterized by the participation ratio
\begin{equation}
R_{n}=(\sum_{i=1}^{N} a_{ni}^2)^2/(N\sum_{i=1}^{N} a_{ni}^4),
\label{5}
\end{equation}%
which is a measure of the portion of the space where the amplitude
of wavefunction differs markedly from zero. For an extended state,
$R$ has a finite value (typically close to 1/3 in the presence of
disorder), whereas for a localized state $R$ approaches zero
proportional to $(1/N)$ \cite{Edward72}. The results for $\xi=1.73a$
with different potential height $V\geq0$ is plotted in Fig.
\ref{F3}. For small $V$ (Fig. \ref{F3} (a) and (b)), where the
electronic behaviors inside and outside the barrier are Dirac-like
(Fig. \ref{F3} (e) and (f)), there are no bound states. When $V$ is
increased, bound states with small $R$ begin to appear in the
negative energy region near the Dirac point, as seen in Fig.
\ref{F3} (c). For positive injected energy (orange arrow with solid
line in Fig. \ref{F3} (g)), the electron is not far from
$\mathbf{K}$ both inside and outside the barrier, so the regime of
Klein tunneling is still valid and the electron cannot be trapped.
On the other hand, for negative energy (orange arrow with dashed
line in Fig. \ref{F3} (g)), the electron sees a non-Dirac barrier
(pointed by the yellow arrow). This causes strong back-scattering
and localization around the impurity. When $V$ is increased further
(Fig. \ref{F3} (h)), even electrons in the positive Dirac region
will encounter strong back-scattering in the barrier and will be
localized (Fig. \ref{F3} (d)). For negative $V$ (not shown here),
all the results are similar, except that the localized states now
first appear in the positive region near Dirac point. In conclusion,
the localized states originate from back-scattering at the barrier
due to its deviation from Dirac behavior at the Fermi level. The
states near the Dirac point with low density of states will be more
sensitive to backscattering and will be localized first, as in the
case of conventional disordered
systems\cite{MacKinnon85,Braun1997,Keith2001}.

Why is the MIT in disordered graphene of K-T type? The K-T
transition is a typical topological transition which has been
understood as unbounding of vortex-anti-vortex pairs
\cite{Kosterlitz73}. For instance, in the high temperature phase 2D
XY model, a plasma of unbounded vortices and anti-vortices of local
spins gives rise to an exponential decay of spin correlation
function; in the low temperature phase, vortices and anti-vortices
are bound to each other, leading to a power law
correlation function. This can be clearly seen in the present
problem if the local currents are identified with local spins in XY
model.

\begin{figure} [t]
\begin{center}
\fbox{\includegraphics[width=0.35\textwidth]{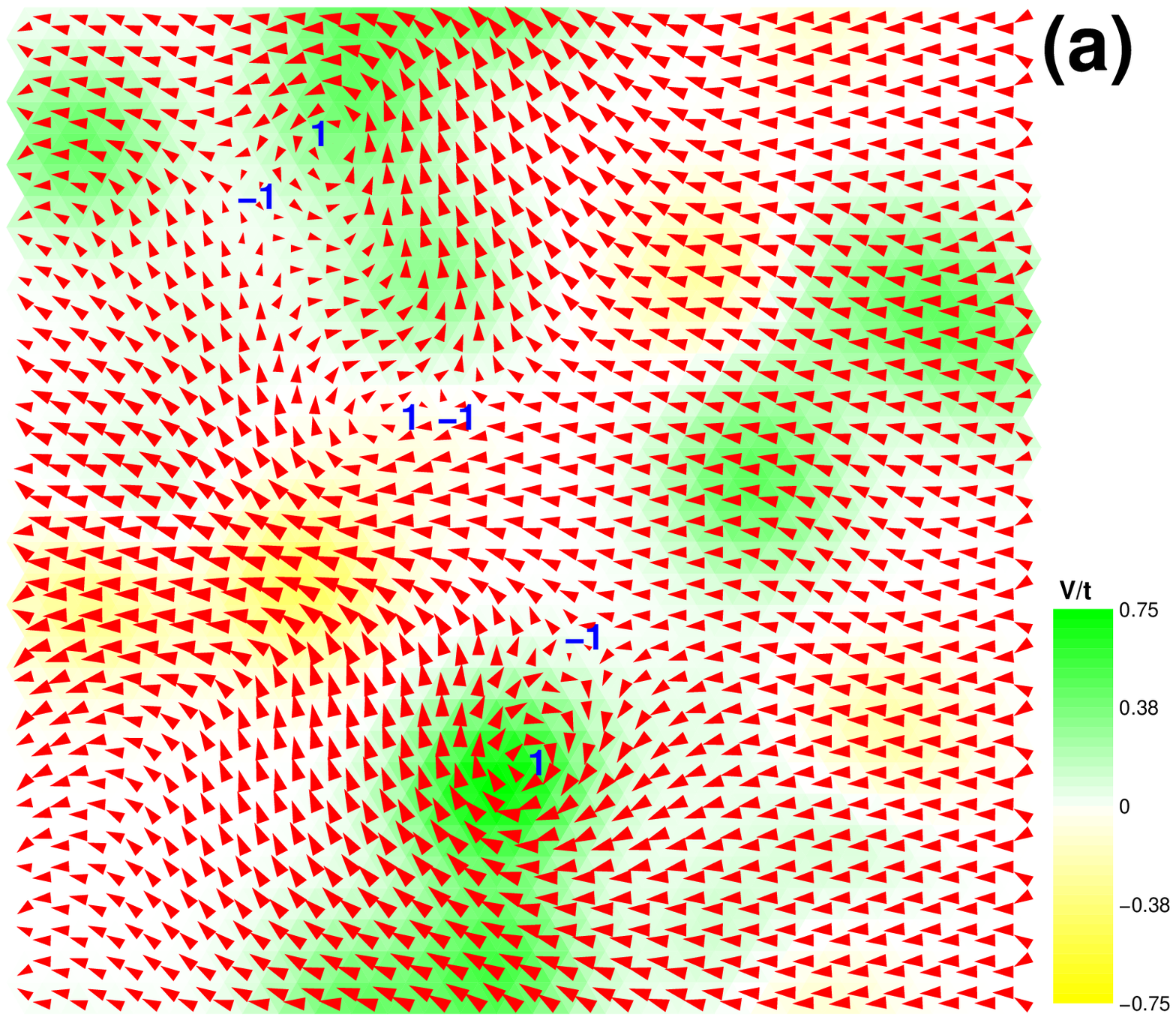}}
\fbox{\includegraphics[width=0.35\textwidth]{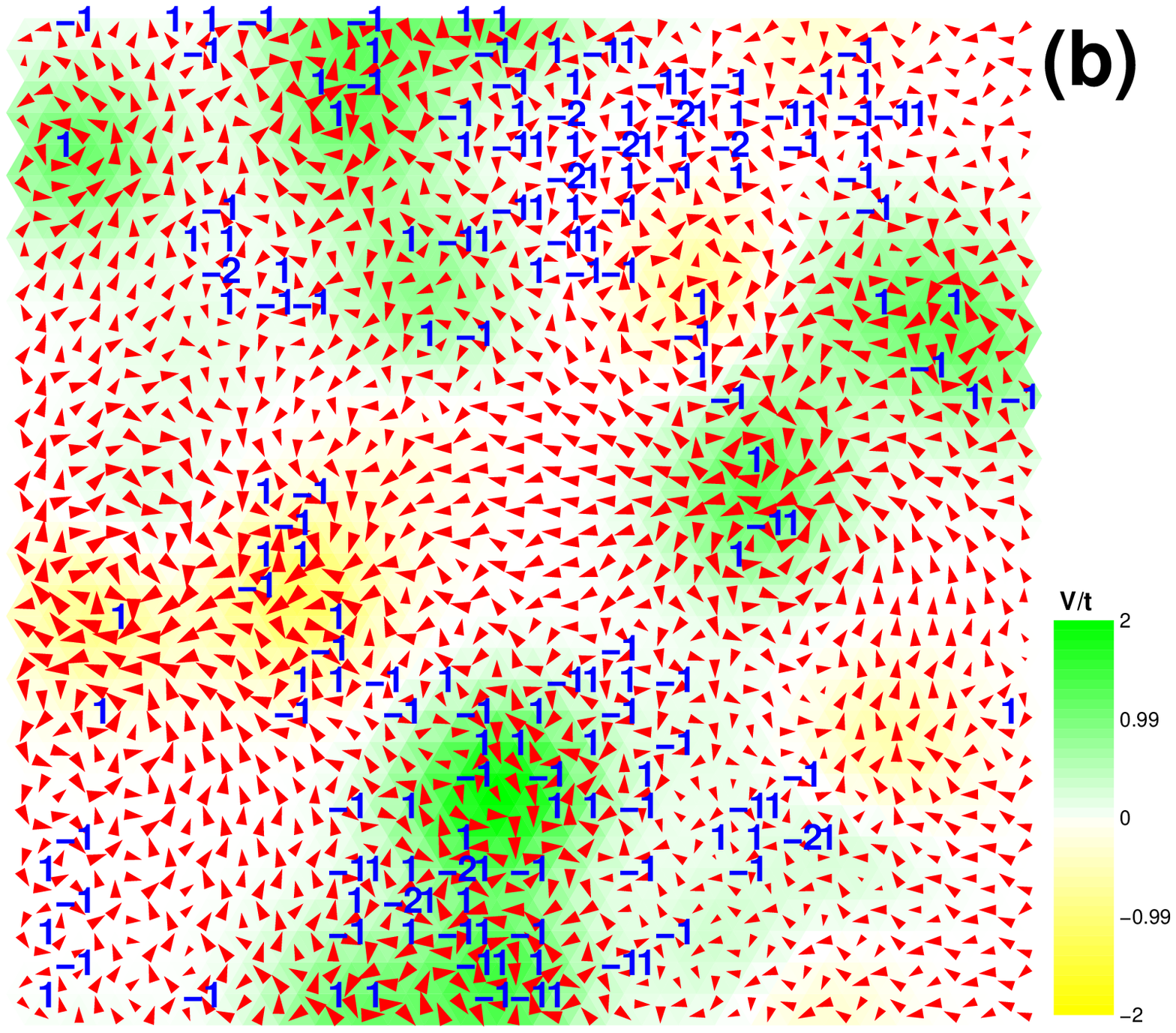}}
\end{center}
\caption{(Color online) Typical configurations of local currents
$\mathbf{i}_{n}$ (red arrows) and potential $V_n$ (color contour) on
two sides of K-T type MIT with $N=56\times32$ sites, $\xi=1.73a$,
$n_I=1\%$ and $E_F=0.1t$. (a): $W=1.1t$ (delocalized); (b): $W=2.9t$
(localized). The size of arrows is proportional to the logarithm of
current value. Carbon hexagons with topological charge $n\neq 0$ are
marked explicitly with blue numbers. Both plots are in the same
random realization of impurities, with different potential height
therefore effectively different $W$. } \label{Vortex}
\end{figure}

The bond current vector $\mathbf{i}_{l\rightarrow m}(E_F)$ per unit
energy pointing along the bond between sites $l$ and $m$ can be
calculated using Green's functions\cite{Datta,Zarbo07,Zhang07}. It
is more convenient to investigate the ``current flow vector''
$\mathbf{i}_{l}=\sum_{m}\mathbf{i}_{l\rightarrow m}$ defined on site
$l$, where the vectorial summation is taken over the nearest
neighbors of site $l$ \cite{Zarbo07}. The current flow
$\mathbf{i}_{l}$ is a vector with angle $\theta_l\in[0,2\pi)$. The
topological charge $n$ of local currents on a closed path can now be
defined as usual: $n=\frac{1}{2\pi}\oint\nabla\theta\cdot d\ell$. In
Fig. \ref{Vortex}, typical distributions of local currents on both
sides of MIT are plotted. As expected from the K-T picture, in the
delocalized phase (Fig. \ref{Vortex} (a)), vortices ($n>0$) and
anti-vortices ($n<0$) are closely bounded, corresponding to the
``low temperature'' phase of 2D XY model with quasi-long range
correlations. In the localized phase (Fig. \ref{Vortex} (b)), there
are a large number of current vortices and anti-vortices. Many of
them are unbounded, corresponding to the ``high temperature'' phase
of 2D XY model without long range correlations. This offers an
explicit picture of the microscopic origin of the disorder driven
K-T transition in graphene.

In conclusion, we find a Kosterlitz-Thouless type metal-to-insulator
transition as a function of disorder strength or Fermi energy in
disordered graphene with strong long-range impurities. We explicitly
demonstrate the KT nature of transition by showing the bounding and
unbounding of local current vortexes. One unique feature about the
K-T transition is a scaling of exponential form near the transition
point. The conductance $g \propto
\exp(-\frac{\alpha}{\sqrt{W-W_c}})$ where $\alpha$ is a constant.
Recently, the MIT near the neutral point of graphene has been
observed in graphene nanoribbons\cite{Adam08} which are
quasi-one-dimensional systems. Our results can be tested in
experiments with large nanoribbon radius.

We thank D.-X. Yao, W.-F. Tsai and C. Fang for useful discussions.
YYZ and JPH were supported by the NSF under grant No. PHY-0603759.

\end{document}